\documentclass[9pt,twocolumn,twoside]{osajnl}

\journal{ol}

\setboolean{shortarticle}{true}

\title{Pulse-regulated single-photon generation via quantum interference in a $\chi^{(2)}$ nonlinear nanocavity}

\author[1]{Yuyi Yan}
\author[1]{Yanbei Cheng}
\author[1]{Shengguo Guan}
\author[1]{Danying Yu}
\author[1,*]{Zhenglu Duan}

\affil[1]{College of Physics and Communication Electronics, Jiangxi Normal University,
Nanchang, 330022, China}
\affil[*]{Corresponding author: duanzhenglu@jxnu.edu.cn}

\begin{abstract}
A scalable on-chip single-photon source at telecommunications wavelengths is an essential
component of quantum communication networks. In this work, we numerically construct a
pulse-regulated single-photon source based on an optical parametric amplifier in a
nanocavity. Under the condition of pulsed excitation, we study the photon
statistics of the source using the Monte Carlo wave-function method. The
results show that there exits an optimum excitation pulse width for
generating high-purity single photons, while the source brightness
 increases monotonically with  increasing  excitation pulse width. More importantly, our system can be operated resonantly and we show that in this case the oscillations in $g^{(2)}(0)$ is completely suppressed.

\end{abstract}

\setboolean{displaycopyright}{true}

\begin{document}

\maketitle

With the rapid development of quantum technologies, quantum information
processing (QIP) has undergone a transition from a scientific research field
to a research-usable technology, with two main practical applications,
namely, quantum communication \cite{QT0} and quantum computation \cite{QC1}.
As reliable information carriers, single photons are indispensable in
applications of photon-based QIP \cite{SPS0}. In particular, a high-quality
single-photon source in telecommunications bands would enable access to
fiber-based quantum communication with low dispersion and low loss \cite%
{SPFiber}.

One possible method for generating single photons is to use the
quantum-interference-induced photon antibunching effect, also known as the
unconventional photon blockade effect \cite{UPB0,UPB}. This photon
preparation method requires only very weak nonlinearity, in contrast to the
strong nonlinearity required to induce conventional photon blockade \cite{PB}%
. Quantum-interference-induced photon antibunching has been proposed
theoretically \cite{UP1} and realized experimentally \cite{UP3,UP4} in
various systems, such as coupled cavities, atomic--optomechanical hybrid
systems and superconducting circuits. Quantum-interference-induced photon
antibunching has also been observed in a degenerate optical parametric
amplifier \cite{QIS1,QIS2}. As pointed out in Ref. \cite{QIS1}, the physical
basis of this mechanism is that destructive interference between a
two-photon transition and a one-photon transition leads to a low or even
vanishing probability of the two-photon state. Since this state is similar
to a coherent field without the two-photon term, it has been termed a
``modified coherent state'' \cite{Modifiedcoherentstate}. More recently, researchers have pointed out \cite{GS} and further clarified \cite{UP1} that this state actually is an optimized Gaussian squeezed state. Such states have been used as single-photon sources
\cite{QIS2} and in quantum cryptography \cite{QKD1}.

In a degenerate parametric down-conversion process in a bulk $\chi ^{( 2) }$
crystal or waveguide, strong second-harmonic light can efficiently generate
fundamental-wave photons under the condition of phase matching. However,
nonlinear interactions between light fields and materials are usually weak,
and hence high light intensities and long interaction times are required to
achieve high conversion efficiency. Fortunately, a high-quality
double-resonance optical cavity can trap both second-harmonic and
fundamental-mode light for a longer time and enhance their effective
intensities, resulting in greatly enhanced nonlinear conversion efficiency
\cite{CSN}. Therefore, devices in which parametric processes occur inside a
microcavity or nanocavity have been proposed and fabricated, such as
photonic crystal cavities \cite{PCS} and micro-ring/disk assemblies \cite%
{MR,MD}, thereby extending traditional bulk optics to the micro- and
nanoscales.

Thanks to the availability of nanofabrication technology, it has been
possible to develop an on-chip single-photon source with strong antibunching
with the advantages of compactness and scalability \cite{SPC}. However, the
challenge remains of preparing an on-demand integrable single-photon source
at telecommunications wavelengths with high purity and efficiency \cite{TS0}%
. To date, most studies of quantum-interference-induced photon antibunching
have dealt with the case of continuous-wave (CW) driving and much less work
has been done on pulsed driving \cite{SPS}, which is required by on-demand
photon sources.

Motivated by the above considerations, in this work, we investigate
single-photon generation under pulsed excitation at telecommunications
wavelengths via quantum interference in a nanocavity made of weak $\chi
^{(2)}$ nonlinear material. For instance, III--V semiconductors, which have
low losses in the near-infrared region, are potential candidates for such $%
\chi ^{(2)}$ nonlinear materials. Compared to those single photon sources
based on strong $\chi^{(2)}$ nonlinearity \cite{X2S} or coupled $\chi^{(2)}$
nonlinear nanocavities \cite{X2C}, the scheme considered in the work merely
requires a single double-resonance nanocavity with weak $\chi^{(2)}$
nonlinearity, which is easier to realize in real experiments. Using the
Monte Carlo wave-function method, we simulate the photon emission process
and study the photon statistics. We find that there is an optimum pulse
width for achieving high purity of the single-photon source and that the
source brightness increases with increasing pulse width. More importantly,
in contrast to the photon antibunching induced in coupled cavity systems
\cite{SPS}, in a certain parameter regime (namely, with zero detuning),
oscillations in $g^{(2)}(\tau )$ are fully suppressed, and therefore our
model is naturally suitable to serve as a pulse-regulated single-photon
source with high purity under pulsed excitation.

\begin{figure}[tbp]
\includegraphics[width=3.2in]{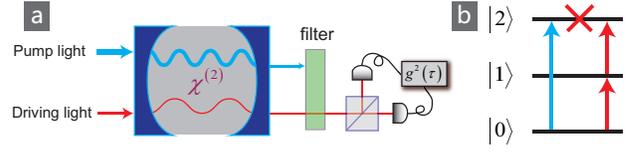}
\caption{(a) Scheme for single-photon generation in a
nonlinear-nanocavity-based optical parametric amplifier driven by a strong
harmonic pump and a weak fundamental driving field. Emitted photons from the
right channel can be collected and analyzed by a photon counter. (b) Energy
level diagram. The quantum destructive interference between different
two-photon events from coherent excitation and parametric down-conversion
leads to a vanishing probability of the two-photon state.}
\label{fig1}
\end{figure}

We consider a model consisting of a driven dissipative nonlinear nanocavity
(i.e., a photonic crystal cavity or a micro-ring/disk), as shown
schematically in Fig. ~\ref{fig1}(a). The fundamental mode and the
second-harmonic mode are spatially overlapping in the same nanocavity at
resonance frequencies $\omega _{a}$ and $\omega _{b}=2\omega _{a}$, respectively. The two
modes are coupled through $\chi ^{(2)}$ nonlinearity that mediates the
conversion of one photon in mode $\hat{b}$ to two photons in mode $\hat{a}$,
and vice versa. The fundamental mode $\hat{a}$ is driven by a weak driving
light with strength $E$ and frequency $\omega _{d}$. The second-harmonic
mode $\hat{b}$ is driven by a strong pump light with strength $F$ and
frequency $2\omega _{d}$. In the rotating frame of the driving and pump lights, the
Hamiltonian of the system is (with $\hbar =1$)
\begin{eqnarray}
\hat{H} &=&\Delta \hat{a}^{\dagger }\hat{a}+2\Delta \hat{b}^{\dagger }\hat{b}%
+\chi \big(\hat{b}\hat{a}^{2\dagger }+\hat{b}^{\dagger }\hat{a}^{2}\big)
\notag \\
&&+E\big(\hat{a}^{\dagger }+\hat{a}\big)+F\big(e^{-i\theta _{0}}\hat{b}%
^{\dagger }+e^{i\theta _{0}}\hat{b}\big),  \label{H0}
\end{eqnarray}%
where $\Delta =\omega _{a}-\omega _{d}$ is the detuning between the
fundamental mode and the driving light, $\chi $ is the
parametric gain, and $\theta _{0}$ is the phase difference between the driving
light and the pump light, which can be adjusted by a movable mirror outside
the nanocavity. In contrast to the system considered in Ref. \cite{UP0}, we
here assume the pump light is very strong and the depletion by the
fundamental mode is negligible, hence the second-harmonic mode can be
approximately expressed as $\hat{b}\simeq F\left/ \sqrt{4\Delta ^{2}+\gamma
^{2}/4}\right. e^{-i\theta }$ with $\theta =\tan^{-1}\left[\gamma /\left(
4\Delta \right) \right]-\theta _{0}$, where we have introduced the decay rate of second-harmonic mode $\gamma$. Substituting this expression into Hamiltonian (\ref{H0}),
we have
\begin{equation}
\hat{H}=\Delta \hat{a}^{\dagger }\hat{a}+E\big(\hat{a}^{\dagger }+\hat{a}%
\big)+U\big(e^{i\theta }\hat{a}^{2\dagger }+e^{-i\theta }\hat{a}^{2}\big),
\label{H}
\end{equation}%
where $U=F\chi/\sqrt{4\Delta ^{2}+\gamma ^{2}/4}$ is the
effective parametric gain. Note that for a coherent driving field, $E$ is
constant, while for pulsed driving, $E$ is time-dependent. The Hamiltonian (%
\ref{H}) describes an optical cavity excited by both parametric and coherent
driving fields, which is the starting point for the following calculation.

According to Ref. \cite{QSSSSSS}, in the limit of weak driving $E\ll \kappa $
and low parametric gain $U\ll \kappa $ (far below the threshold for
parametric oscillation), one can obtain the optimum condition for strong
antibunching: $E^{2}=U\sqrt{\Delta ^{2}+\kappa ^{2}/4}$ and $\allowbreak
\theta =\tan ^{-1}( \kappa /2\Delta ) $. The strong antibunching comes from
the quantum destructive interference between two different two-photon
events: one coming from coherent driving, $\vert 0\rangle \overset{E}{%
\longrightarrow }\vert 1\rangle \overset{E}{\longrightarrow }\vert 2\rangle $%
, and the other from parametric down-conversion, $\vert 0\rangle \overset{U}{%
\longrightarrow }\vert 2\rangle $. Such a system can be viewed as a
stochastic single-photon source under CW driving. However, for a
deterministic single-photon source, one requires single-photon emission at
determined times. Therefore, in order to prepare a pulse-regulated
single-photon source, our system should be operated in the pulsed regime.

Given that single-photon sources in the telecommunications band are of great
interest in the context of quantum information processing on chips, the
operational wavelength of the fundamental mode of the nanocavity is assumed
to be 1.5\,$\mu$m. To date, various nanocavities made of noncentrosymmetric
materials (e.g., GaP and GaAs) with high quality factor ($Q=10^{5}\thicksim10^{6}$%
) have been designed and fabricated \cite{HQ0}. Therefore, we can choose a
nanocavity with a quality factor $Q\sim 10^{6}$, which gives a cavity-mode
linewidth $\kappa \sim 1$\,GHz. For III--V semiconductor materials, the
second-order susceptibility $\chi ^{(2)}$ is typically of the order of $%
10^{-10}$\,m/V. If the nanocavity is made up of photonic crystal cavities,
for an optimum geometric configuration, the realistic maximum value of the
nonlinear coupling $\chi $ can reach $1$\,GHz. Hence we can safely assume a
suitable parametric gain $U$ in the following study.

To study the photon statistics of the emitted field in the pulsed-driving
situation, we need to analyze the second-order correlation function with
delay. Using the quantum regression theorem \cite{QN}, we have the
time-dependent correlation function
\begin{align}
g^{(2)}( \tau ) &=\frac{\big\langle \hat{a}^{\dagger }\hat{a} ^{\dagger }(
\tau ) \hat{a}( \tau ) \hat{a} \big\rangle }{\big\langle \hat{a}^{\dagger }%
\hat{a}\big\rangle ^{2}}  \notag \\[3pt]
&=\frac{\mathrm{Tr}\,\big\{ \hat{a}^{\dagger }\hat{a}e^{\hat{L}\tau }\big[
\hat{a}\hat{\rho} _{ss}\hat{a}^{\dagger }\big] \big\} }{\mathrm{Tr}\,\big(
\hat{a}^{\dagger }\hat{a}\hat{\rho} _{ss}\big) ^{2}},  \label{g2}
\end{align}%
where $\hat{\rho}_{ss}$ is the reduced density matrix for mode $\hat{a}$ in
the steady state, and the superoperator $\hat{L}$ is defined as
\begin{equation}
\hat{L}\hat{\rho}=-i\left[ H,\hat{\rho}\right] +\frac{\kappa }{2}\hat{D} %
\left[ \hat{a}\right] \hat{\rho},  \label{L}
\end{equation}%
with the Lindblad operator $\hat{D}\,\big[ \hat{A}\big] \,\hat{\rho}=2\hat{A}%
\hat{ \rho}\hat{A}^{\dagger }-\hat{A}^{\dagger }\hat{A}\hat{\rho}-\hat{\rho}%
\hat{A} ^{\dagger }\hat{A}$.

\begin{figure}[tbp]
\includegraphics[width=3.2in]{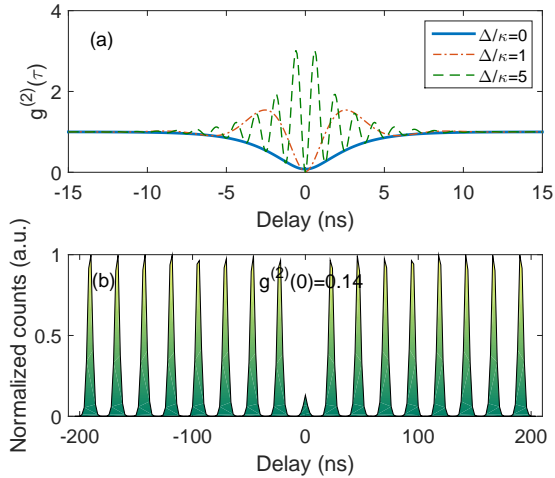}
\caption{(a) Second-order correlation function as a function of time delay $%
\protect\tau $ with different detunings $\Delta $. The driving strength $%
E=50 $\,MHz. (b) Normalized coincident counts with pulsed excitation. The
equal-time second-order correlation function $g^{(2)}(0) =0.14$. The width
of the excitation pulse is 2\,ns, the interval is 24\,ns, and the maximum
amplitude of the driving strength $E_{0}=50$\,MHz.}
\label{fig2}
\end{figure}

Figure~\ref{fig2}(a) shows the second-order correlation function as a
function of time delay for different detunings. One can see that for nonzero
detuning, the second-order correlation function oscillates with period $2\pi
/\Delta $, and the output light is antibunched over a time delay shorter
than $1/\Delta $. Similar oscillatory phenomena in $g^{(2)}(\tau )$ have
been observed in other situations \cite{delay1,UPB}. This behavior implies
that the width of the driving pulse $\Delta t$ should be larger than $%
1/\Delta $ to guarantee strong antibunching of the output light. However, in
the case of zero detuning, $g^{2}( \tau ) <1$ for all time delays. This
suggests that the limitation on the width of the driving pulse no longer
exists in the zero-detuning case of our model. Therefore, our model is well
suited for generating photon antibunching with pulsed excitation.

Next we turn to demonstrating the process of antibunching photons emitted
from the nanocavity driven by a series of light pulses. In our numerical
calculation, the driving field is assumed to be $E=E_{0}\exp \left[ -\big(
t-nt_{0}\big) ^{2}/\Delta t^{2}\right] $, with $\Delta t$ being the pulse
width and $t_{0}$ the time at which the driving reaches its maximum value $%
E_{0}$. To achieve ideal antibunching in the case of pulsed excitation, the
parametric gain and relative phase have to satisfy optimum conditions.
Obviously, the parametric gain is also time-dependent, i.e., $U( t) =E^{2}(
t) /\sqrt{\Delta ^{2}+\kappa ^{2}/4}$, which requires the pump light to have the same temporal shape as the excitation light $E^{2}$. To avoid
overlapping of adjacent excitation pulses, we set the interval between them
as 12 times the pulse width. Of course, there are constraints on the pulse
width. For instance, the spectral width of the pulses should be smaller than
the nonlinear shift of the energy levels.

To better illustrate the quantum statistics of the output light, we adopt
the Monte Carlo wave-function method to simulate the stochastic evolution of
the system and count the photon number from the output port \cite{MC}, which
closely mimics the Hanbury Brown--Twiss (HBT) experimental procedure. We
extract the correlation function information from a single but sufficient
long trajectory. Below, we briefly introduce the procedure involved in using
the Monte Carlo wave-function method to simulate the stochastic photon
emission from the nanocavity in our model.

First, we choose a pure state $\vert \psi ( t) \rangle $ as the system
state, and then let it evolve nonunitarily for a very short but finite time
interval $\delta t$ ($\kappa\delta t \ll 1$) as
\begin{equation}
\vert\tilde{\psi}( t+\delta t) \rangle =\left( 1- \frac{iH_\mathrm{eff}%
\delta t}{\hbar }\right) \vert \psi ( t) \rangle,
\end{equation}
where the non-Hermitian Hamiltonian $\hat{H}_\mathrm{eff}=\hat{H}-i\hat{J}%
^{\dag } \hat{J}/2$, with $\hat{J}=\sqrt{\kappa }\hat{a}$ being the jump
operator. The second term in $\hat{H}_\mathrm{eff}$ is non-Hermitian,
representing the dissipation induced by coupling with the environment. The
norm of the evolved function wave $\langle \tilde{\psi}( t+\delta t)
\,\vert\, \tilde{\psi}( t+\delta t) \rangle =1-\delta p$ can be considered
as the probability of no photon emission happened in the time interval $%
\delta t$, and $\delta p=$ $\delta t\langle \tilde{\psi}( t) \vert\, \hat{J}%
^{\dagger }\hat{J}\,\vert \tilde{\psi}( t) \rangle $ corresponds to the
probability of a photon emission event.

Second, we decide whether an emission event occurs or not by comparing $%
\delta p$ with a uniformly generated random number $r$ ($0\leq r\leq 1$). If
$r>\delta p$, no photon emission event occurs, and we normalize the wave
function:
\begin{equation}
\vert \psi ( t+\delta t) \rangle =\frac{\vert \tilde{\psi}( t+\delta t)
\rangle }{\sqrt{1-\delta p}}.
\end{equation}%
Otherwise, one photon is emitted, and the system state is collapsed to a new
state:
\begin{equation}
\vert \psi ( t+\delta t) \rangle =\frac{\hat{J}\, \vert \tilde{\psi}(
t+\delta t) \rangle }{\sqrt{\delta p/\delta t}}.
\end{equation}%
The state $\vert \psi ( t+\delta t) \rangle $ will serve as the initial
state for the next step in the iteration. By repeating this procedure, we
can obtain the wave function of the system state at all times, and the
system properties, including the second-order correlation function of the
photons, can be evaluated based on the wave function.

We now show how to extract the correlation function from the simulated
results based on the Monte Carlo wave function. Assuming a photon count
taking place at time $t$, we cumulate the emitted photon number $N( t,t+\tau
) $ in an interval $\Delta \tau $ at $t+\tau $. To achieve a balance between
resolution and statistical fluctuations, the interval $\Delta \tau $ should
be carefully chosen in the calculation. We average the $N( t,t_{j}) $ over
all $t$ to obtain $\bar{N}( t,t_{j}) =\sum_{t}N( t,t_{j}) /N_\mathrm{total}$%
, where $N_\mathrm{total}$ is the total number of photons emitted during the
total counting duration $T $. Statistically, for a sufficiently large
sample, this mean value $\bar{N}$ is closely related to the conditional
probability of finding a second photon at time $t+\tau $ provided that a
first photon was detected at time $t$. Finally, the second-order correlation
function with time delay is found as
\begin{equation}
g^{(2)}( \tau ) =\frac{\bar{N}( t,t+\tau ) }{\bar{N} ( t) },  \label{g2t}
\end{equation}%
where $\bar{N}( t) =N_\mathrm{total}\Delta \tau /T$ is the mean photon
number in an interval $\Delta \tau $ in the steady state. In the simulation,
we set the total number of excitation pulses as $10^{7}$, and the width and
amplitude of each excitation pulse are $2$\,ns and $0.05$\,eV, respectively.
The detuning $\Delta =0$ and the interval between excitation pulses is $24$%
\,ns, corresponding to a $42$\,MHz repetition rate. Figure~\ref{fig2}(b)
shows the normalized number of counts versus the time delay. The
second-order correlation function $g^{(2)}( 0) \simeq 0.14$ is calculated
from the integrated number of photon counts in the zero-delay peak divided
by those in its adjacent peak. In the numerical experiment, we observe a
count rate of $800\,000$ per second, under excitation at a repetition rate
of $42$\,MHz, which gives an overall system efficiency of $1.9$\%.

\begin{figure}[tbp]
\includegraphics[width=3.2in]{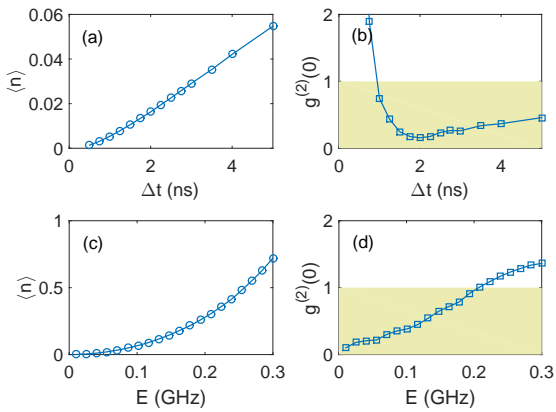}
\caption{(a) Mean photon number (brightness) and (b) equal-time second-order
correlation function (purity) as functions of excitation pulse width with
maximum driving strength $E_{0}=50$\,MHz. (c) Mean photon number
(brightness) and (d) equal-time second-order correlation function (purity)
as functions of the maximum driving strength $E_{0}$ for pulse width $\Delta
t =2$\,ns. }
\label{fig4}
\end{figure}

From a practical viewpoint, the purity and brightness of a single-photon
source are two important features of merit. Here we investigate the
influence of the parameters of the excitation pulse upon these features. As
described in the literature, the purity is usually characterized by the
zero-delay second-order correlation function $g^{(2)}(0)$ and the brightness
by the emitted photon number per excitation pulse $\langle n\rangle $ \cite%
{purity}. Figure~\ref{fig4} shows numerical results for the brightness and
purity as functions of the width and strength of the excitation pulse. For
fixed maximum strength $E_{0}=50$\,MHz [as shown in Fig.~\ref{fig4}(a)], the
brightness increases monotonically with increasing width $\Delta t$. Figure~%
\ref{fig4}(b) shows that $g^{(2)}(0)$ decreases and then increases with
increasing width $\Delta t$. The minimum value of $g^{(2)}(0)$ can reach $%
0.14$, corresponding to high purity of the single-photon source. The reason
for this behavior is that a small temporal width of the excitation pulse
corresponds to a large width in the frequency domain, and consequently most
frequency components of the pulse deviate dramatically from the optimum
condition, leading to an increase in the second-order correlation function.
On the other hand, longer pulse excitation means an increased probability
for the detector to collect two or more photons per excitation pulse, and
this also increases $g^{(2)}(0)$. Therefore, an optimum width for the
excitation pulse is helpful to achieve a minimum $g^{(2)}(0)$. Figures~\ref%
{fig4}(c) and \ref{fig4}(d) illustrate that, for a fixed width of the
excitation pulse $\Delta t=2$\,ns, both brightness and $g^{(2)}(0)$ increase
monotonically as the maximum driving strength is increased. This can be
attributed to an increase in the population of multiphoton states brought
about by the increased driving strength.

In conclusion, we have investigated pulse-regulated integrable single-photon
generation in a weak $\chi ^{(2)}$ nonlinear nanocavity and have analyzed
statistical properties of the emitted photons via the Monte Carlo
wave-function method. For a typical cavity mode linewidth $\kappa \sim 1$%
\,GHz, the maximum single-photon repetition rate reaches $42$\,MHz, the
purity is $\sim$$0.14$, and the efficiency is $1.9$\%. We have found that
for fixed driving strength, there exists an optimum pulse width for
achieving maximum purity and that the purity can be further improved at the
cost of reduced generation efficiency by decreasing the driving strength.
Our work may offer direct guidance for experimental construction of
pulse-regulated single-photon sources via quantum interference.

\section{Funding Information}

National Natural Science Foundation of China (NSFC) (11664014, 11504145, and
11364021); Natural Science Foundation of Jiangxi Province (20161BAB201023
and 20161BAB211013).

\bibliographystyle{plain}
\bibliography{Reference}

\bibliographyfullrefs{Reference}

\end{document}